\documentclass[conference]{IEEEtran}
\IEEEoverridecommandlockouts
\pdfoutput=1

\usepackage{cite}
\usepackage{amsmath,amssymb,amsfonts,amsthm}
\usepackage{graphicx}
\usepackage{textcomp}
\usepackage{xcolor}
\usepackage{balance}
\usepackage{geometry}                
\usepackage{graphicx}
\usepackage{amssymb}
\usepackage{epstopdf}
\usepackage[all]{xy}
\usepackage{threeparttable}
\usepackage{color}
\usepackage[linesnumbered,ruled,lined]{algorithm2e}
\usepackage{algpseudocode}
\usepackage{amsxtra}
\usepackage{amscd}

\numberwithin{table}{section}
\newtheorem{thm}{Theorem}[section]

\newtheorem{lem}[thm]{Lemma}
\newtheorem{rem}[thm]{Remark}

\def\BibTeX{{\rm B\kern-.05em{\sc i\kern-.025em b}\kern-.08em
    T\kern-.1667em\lower.7ex\hbox{E}\kern-.125emX}}

\begin{document}

 \title{Three-Party Secure Protocols for G-module and their Applications}

\author{\IEEEauthorblockN{1\textsuperscript{st} Qizhi Zhang}
\IEEEauthorblockA{\textit{Ant Group} \\
Hangzhou, China \\
qizhi.zqz@antgroup.com}
\and
\IEEEauthorblockN{2\textsuperscript{nd} Lichun Li}
\IEEEauthorblockA{\textit{Ant Group} \\
Hangzhou, China \\
lichun.llc@antgroup.com}
\and
\IEEEauthorblockN{3\textsuperscript{rd} Shan Yin}
\IEEEauthorblockA{\textit{Ant Group} \\
Hangzhou, China \\
yinshan.ys@antgroup.com}
\and
\IEEEauthorblockN{4\textsuperscript{th} Juanjuan Sun}
\IEEEauthorblockA{\textit{School of mathematical sciences, Tongji university} \\
Shanghai, China \\
Corresponding author\\
sunjuan@tongji.edu.cn}
}

\maketitle

\begin{abstract}

Secure comparison and secure selection
are two fundamental MPC (secure Multi-Party Computation)
protocols. One important application of
these protocols is the secure ReLU and DReLU
computation in privacy preserving deep learning. In
this paper, we introduce G-module, a mathematics tool,
to re-design such protocols. In mathematics, given a group G, 
a G-module is an abelian group M on which G acts compatibly with the abelian group structure on M.
 We design
three secure protocols for three G-module operations. i.e. ``G-module action", ``shared G-module action" and ``G-module recover".
As far as we know, this is the first work on secure G-module
operations. Based on them, we design secure
comparison, selection, ReLU and DReLU protocols,
which improve communication efficiency by 2X to 10X
compared with state of arts. Our protocols are very
computation efficient too. They do not require public
key operations or any other expensive operations.
\end{abstract}

\begin{IEEEkeywords}
MPC, G-module, Secure Compare, ReLU
\end{IEEEkeywords}

\section{Introduction}

Secure multi-party computation (MPC) is a subfield
of cryptography with the goal of creating
method for multiple parties to jointly compute a
function over their inputs while keeping those inputs
privately. Secure comparison and secure selection
are two important and fundamental MPC
protocols widely used in MPC applications. One
important application is Privacy Preserving Deep
Learning (PPDL), which allows a group of independent
data owners to collaboratively learn a
neural network model over their data sets without
exposing their private data. Secure comparison and
secure selection are the major building blocks for
secure computing ReLU and DReLU, two important
functions in the training of neural network.
In this paper, we study secure comparison and
selection protocols as well as close related secure
ReLU and DReLU protocols.

%

\subsection{Related  Works}

 Secure Comparison (Yao's Millionaires' problem) is an important and  classical problem in MPC. It has been widely studied, for example, in \cite{Yao}, \cite{similar}, \cite{SC29}, \cite{SC32}, \cite{SC62}, \cite{FSS}, \cite{GSV07}, \cite{KSS09}  and so on. The protocol in \cite{FSS} is the state of art
 of  Secure Comparison.
 There is a close connection between Secure Comparison  and DReLU.
 We will give this connection in Lemma \ref{SCtoSReLU} precisely.

ReLU is the necessary layer in deep learning. The MPC protocols of DReLU and ReLU is wildly studied in, for example, 
\cite{ReLU49}, \cite{ReLU50}, \cite{ReLU51}, \cite{ReLU56}, \cite{secureNN}, \cite{CrypTFlow}.
In SecureNN \cite{secureNN}, 3-party protocols to compute DReLU and ReLU are given.
The main tools 
are Private Compare and Select Share. The functionality dependence is shown in Figure \ref{SecureNN}. In CrypTFlow \cite{CrypTFlow}, PRF (Pseudo-Random Function) is used to reduce the communication in  SecureNN. 
The improved protocols $\prod _\textbf{DReLU} (\{P_0, P_1\}, P_2)$ and $\prod _\textbf{ReLU} (\{P_0, P_1\}, P_2)$ are the  states of art for the functionalities $\mathcal{F}_{DReLU}$ and $\mathcal{F}_{ReLU}$ respectively.

 The protocol $\prod _\textbf{SS} (\{P_0, P_1\}, P_2)$ in SecureNN \cite{secureNN} is the state of art of the 3-party protocol realizing the functionality Select Share.

\subsection{Our Contributions}
In this paper, we make four main contributions:

\begin{itemize}

\item[-] Define three new functionalities ``G-module Action", ``shared G-module Action", ``G-module Recover"  in secure multiparty computation for  the  mathematic object: G-module \cite{GTM-4}.  And give 3-party secure protocols realizing these functionalities.

\item[-] As an application of the protocol ``G-module Recover", we give a  new 3-party protocol securely realizing the functionality Secure Comparison (i.e, the Millionaire's Problem), whose total communication (offline+online) is {\color{red}less} than the state of art  about {\color{red}$90\%$} (Table \ref{T_Compare_SC}).

\item[-] We give a protocol  securely realizing the functionality DReLU. 
The online communication, total (online+offline) communication and the round of our DReLU  protocol are {\color{red}less} than the state of art about {\color{red}$69\%, 45\%$ and $60\%$} respectively in the usual setting of size of numbers (Table \ref{T_Compare_DReLU})

\item[-] As an application of the protocol ``shared G-module Action", we give a new 3-party protocol securely realizing the functionality Select Share. 
The online communication, total (online+offline) communication and the round of our Select Share protocol  is {\color{red}less} than
the state of art about  {\color{red}$59\%, 38\%$ and $50\%$} respectively in the usual setting of size of numbers (Table \ref{T_SSN}).

\item[-] We give a protocol securely realizing the functionality ReLU. 
The online communication, total (online+offline) communication and the round of our ReLU protocol are {\color{red}less} than the sate
of art about  
{\color{red}$69\%$, $45\%$ and $60\%$}
 in the usual setting of size of numbers (Table \ref{T_Compare_ReLU}).

\end{itemize}

Moreover, our protocols $\prod _\text{DReLU} (\{P_0, P_1\}, P_2)$ and $\prod _\text{ReLU} (\{P_0, P_1\}, P_2)$ realize the functionalities $\mathcal{F}_{DReLU}$  and $\mathcal{F}_{ReLU}$ on the whole domain $\left[ 0, 2^n-1 \right]$, while 
  the protocols DReLU, ReLU protocol in \cite{secureNN} and \cite{CrypTFlow} just realize the functionalities $\mathcal{F}_{DReLU}$  and $\mathcal{F}_{ReLU}$ on a subdomain $\left[0, 2^k \right] \cup \left[2^n-2^k, 2^n \right)$ of $\left[ 0, 2^n-1 \right]$, where $k<n-1$.

Besides our protocol is secure under the commodity model \cite{Taas}, since the assistant third party $P_2$ only sends message to the parties $P_0$, $P_1$ in offline phase. In online phase, just $P_0$ and $P_1$ play the protocol. The assistant third party $P_2$ does not need to receive any message from the $P_0$, $P_1$ at all. The security of our protocol are much better than that in \cite{secureNN} and  \cite{CrypTFlow}, where the assistant third party needs to both send and receive message online.

\subsection{Our Techniques}

\begin{figure}[htbp]
  \begin{minipage}[t]{0.4\linewidth}
 \includegraphics[scale=0.2]{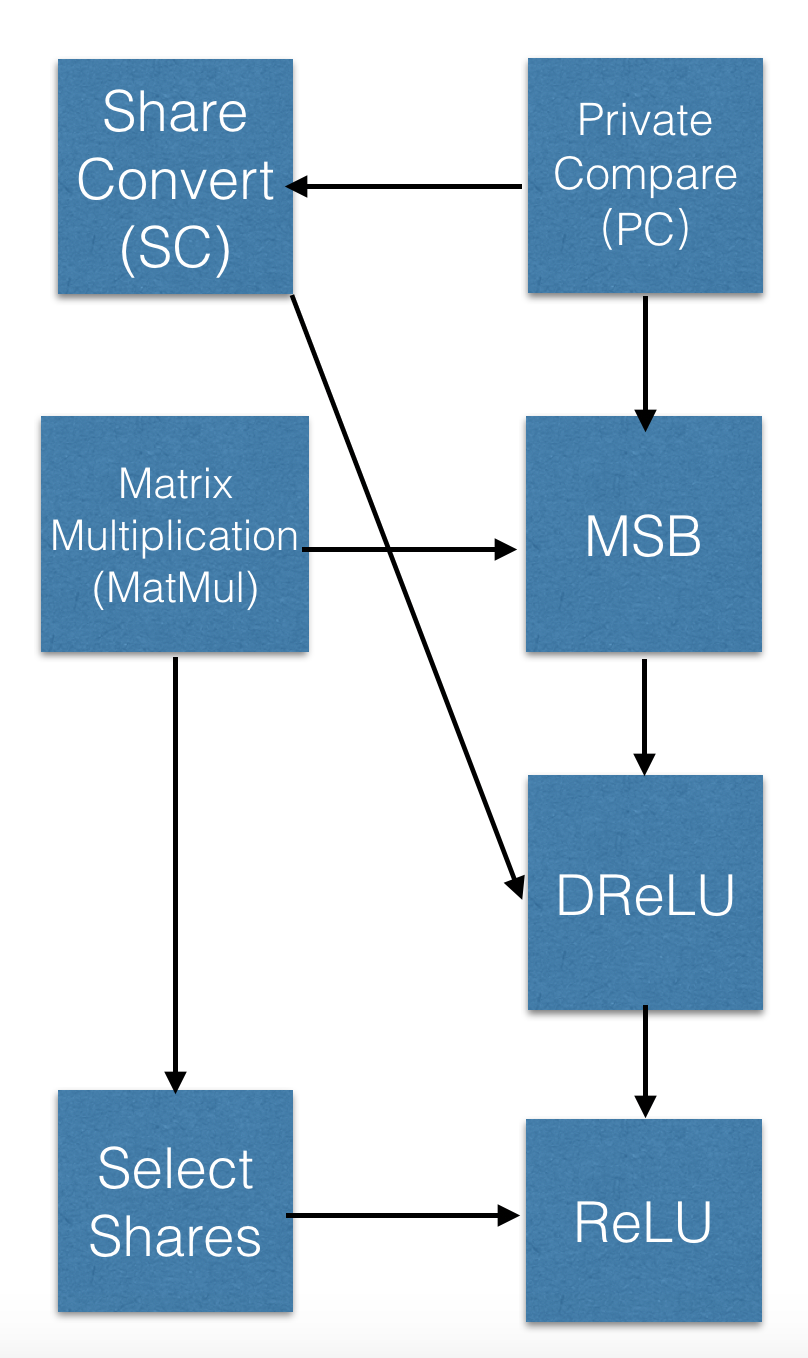}
\caption{Functionality dependence of protocols in SecureNN}
\label{SecureNN}
\end{minipage}
\begin{minipage}[t]{0.3\linewidth}
  \includegraphics[scale=0.25]{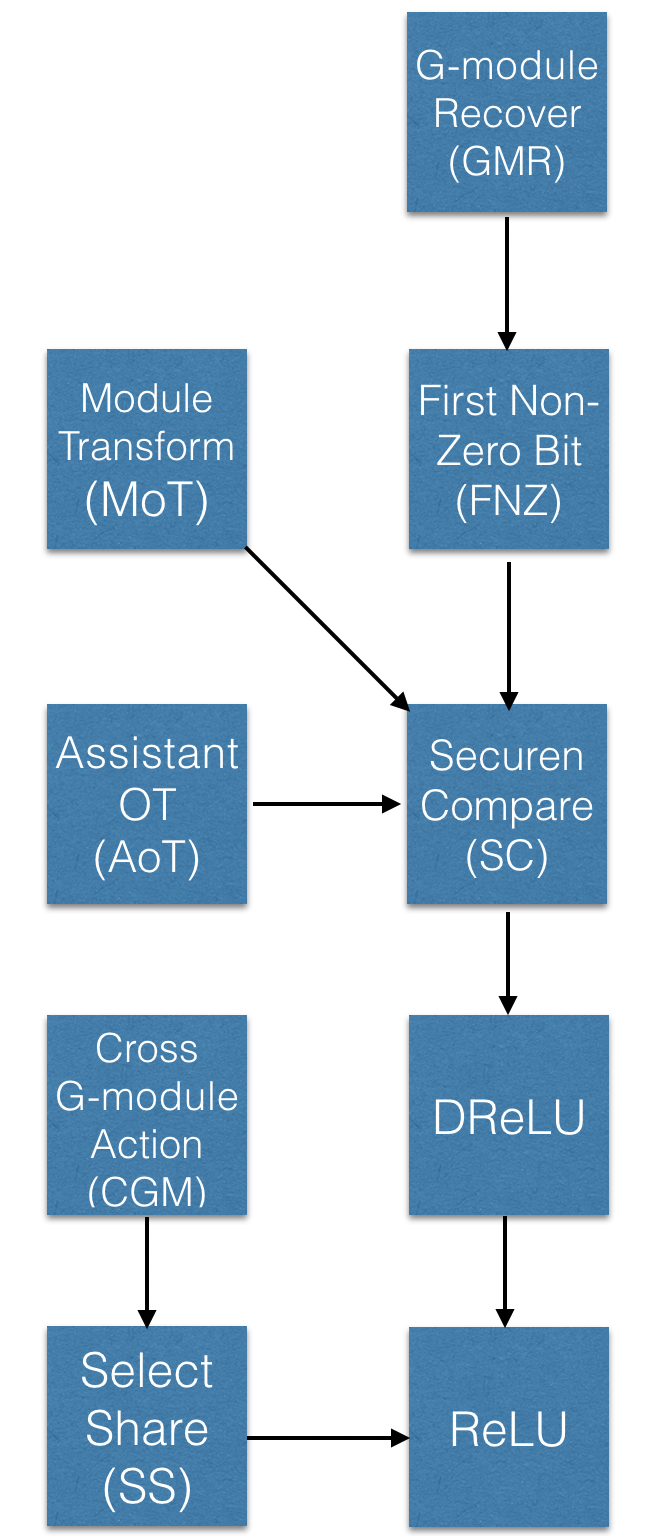}
\caption{Functionality dependence of protocols in this paper}
\label{my}
\end{minipage}
\end{figure}

In our paper, the functionality dependence of protocols  is given in Figure \ref{my}. 
We define a functionality ``G-module Recover" $\mathcal{F}_{GMR}$ , and give a protocol $\prod _\textbf{GMR} (\{P_0, P_1\}, P_2)$ to securely realize it. We give a protocol $\prod _ \textbf{FNZ} (\{P_0, P_1\}, P_2)$ to securely realize the functionality ``First Non-Zero bit" $\mathcal{F}_{FNZ}$ in the $\mathcal{F}_{GMR}$-hybrid model.  
Then we give a protocol $\prod _\textbf{SC} (\{P_0, P_1\}, P_2)$ to securely realize the functionality Secure Comparison $\mathcal{F}_{SC}$ in the
($\mathcal{F}_{FNZ}, \mathcal{F}_{MoT}, \mathcal{F}_{AOT}$)-hybrid model, where the functionality ``Module Transform" $\mathcal{F}_{MoT}$ can be securely realized by the protocol $\prod _\textbf{MoT} (\{P_0, P_1\}, P_2)$ in  \cite{similar}, and
the functionality ``Assistant OT"  $\mathcal{F}_{AOT}$ can be securely realized by the protocol $\prod _\textbf{AOT} (\{P_0, P_1\}, P_2)$ in section \ref{S_AOT}. 
Our protocol $\prod _\textbf{SC} (\{P_0, P_1\}, P_2)$ has the minimal total communication
(offline+online) compare to the state of art (Table \ref{T_Compare_SC}).

Then we  give our protocol $\prod _ \text{DReLU} (\{P_0, P_1\}, P_2)$ to  securely realize the functionality $\mathcal{F}_{DReLU}$
in the $\mathcal{F}_{SC}$-hybrid model. Comparing to the
DReLU protocols  in  SecureNN (\cite{secureNN}) and CrypTFlow (\cite{CrypTFlow}), our protocol has smaller online communication, round and total (online+offline) communication (Table \ref{T_Compare_DReLU}) .

On the other hand, we define a functionality ``shared G-module action" $\mathcal{F}_{SGM}$, and give a protocol $\prod _\textbf{SGM} (\{P_0, P_1\}, P_2)$ to securely realize it.  Then we give a protocol $\prod _\textbf{SS} (\{P_0, P_1\}, P_2)$ to securely realize the  functionality Select Share $\mathcal{F}_{SS}$ in the $\mathcal{F}_{SGM}$-hybrid model. Comparing to the
Select Share protocol in  SecureNN (\cite{secureNN}), our protocol $\prod _\text{DReLU} (\{P_0, P_1\}, P_2)$ has less online communication, round and total (online+offline) communication (Table \ref{T_SSN}).


Finally, we give a protocol $\prod _\text{ReLU} (\{P_0, P_1\}, P_2)$ to  securely realize the functionality 
ReLU $\mathcal{F}_{ReLU}$ in the $(\mathcal{F}_{DReLU}, \mathcal{F}_{SS})$-hybrid model.
Comparing with the ReLU protocol in \cite{secureNN} and \cite{CrypTFlow}, our  protocols $\prod _\text{ReLU} (\{P_0, P_1\}, P_2)$
 have smaller online communication, round and total (online+offline) communication (Table \ref{T_Compare_ReLU}).

\subsection{Organisation}
We  give the functionalities ``G-module Action", ``shared G-module Action", ``G-module recover" and the 3-party
protocols  $\prod _\textbf{GM} (\{P_0, P_1\}, P_2)$, $\prod _\textbf{SGM} (\{P_0, P_1\}, P_2)$, $\prod _\textbf{GMR} (\{P_0, P_1\}, P_2)$  to realize them in Section \ref{S_G_module}. 

We give a 3-party protocol $\prod _\textbf{AOT} (\{P_0, P_1\}, P_2)$ realizing the functionality Assistant OT in 
Section \ref{S_AOT}. We give a 3-party protocol  $\prod _\textbf{MoT} (\{P_0, P_1\}, P_2)$ to realize the functionality Module Transform in Section \ref{S_MoT}. 

We give  a 3-party protocol  $\prod _\textbf{SC} (\{P_0, P_1\}, P_2)$ to realize the functionality
Secure Comparison  in  Section \ref{S_SC}.

We give a 3-party protocol $\prod _\textbf{DReLU} (\{P_0, P_1\}, P_2)$ to realize the functionality
DReLU in Section \ref{S_DReLU}.  

We give a 3-party protocol $\prod _\textbf{SS} (\{P_0, P_1\}, P_2)$ to realize the functionality
Select Share in Section \ref{S_SS}. 

We give a 3-party protocol $\prod _\textbf{ReLU} (\{P_0, P_1\}, P_2)$ to realize the functionality 
ReLU in Section \ref{S_ReLU}.

\subsection{Notation and Terminology}

Let us introduce some notations and  terminologies using in this paper:

$\mathbb{Z}$: The ring of integral numbers;

$\mathbb{Z}/N\mathbb{Z}$: The residue class ring of $\mathbb{Z}$ module $N\mathbb{Z}$;

$\mathbb{F}_q$: The finite field of $q$ elements.

Let (A, +) be an abelian group, for an element $x$ in A, we call $(x_L, x_R) \in A^2$ the shares (or share representation) of $x$ over $A$ , if $x=x_L+x_R \in A$.

If $P_0$, $P_1$ are two parties, we say $\lq\lq P_0$, $P_1$ hold the shares (or share representation) of $x$ over A", if $P_0$ holds an element $x_L\in A$, $P_1$ holds an element $x_R\in A$ such that $x_L+x_R=x \in A$.

Let $G$ be a group with identity element $1_G$ and let $M$ be an abelian group, we call $M$ a G-module (\cite{G-module}, p.186 in \cite{GTM-4}) if there is a map
\[
\begin{array}{rclcc}
G & \times  & M  & \longrightarrow & M \\
(g &,& m) & \longmapsto & gm
\end{array}
\]
satisfying the following properties:

a. For any $m \in M$, one has $1_G m =m$;

b. For any $g_1, g_2 \in G$ and any $m \in M$, one has $(g_1g_2) m = g_1(g_2m)$;

c. For any $g \in G$ and any $m_1, m_2 \in M$, one has $g(m_1+m_2) = gm_1 + gm_2$.

\begin{table*}[htbp]
\caption{Communication of Protocols}
\label{T_all_protocols}
\begin{center}
    \begin{threeparttable}
\begin{tabular}{c|c|c|c|c}
\hline \hline
Protocol & offline com.  & online comm.  & online  round & total comm.\\ \hline \hline
GM(G,A) &  $\log |A|$  & $\log |G|+ \log |A|$ & 1 & $\log |G|+ 2\log |A|$\\    \hline
SGM(G,A) &  $\log |A|$  & $2\log |G|+ 2\log |A|$ & 1 & $2\log |G|+ 3\log |A|$\\    \hline
GMR(G,A) &  $\log |A|$  & $2\log |A|$ & 2 & $3\log |A|$\\    \hline \hline
AOT(A,B) &  $ \log |B|$  & $|A|\log |B|+ \log |A|$ & 2 & $  |A| \log |B| + \log |A| + \log |B|$\\    \hline  \hline
MoT(m) &  $\log m$  & $2$ & 1 & $ \log m+2 $\\ \hline  \hline
FNZ(p, n) & $n \log p \tnote{**} $  & $2 n\log p$ & 2 & $3n \log p$\\    
SC(n) & $ (2n+1)\log p  \tnote{***} \atop +1 $ & $2(n+1)\log p + 3n  \atop + \log (n+1) +1$ & 5 &  $(4n+3)\log p+3n \atop +\log (n+1)+2$  \\
SC(n) (reduced) & $ (2n+1)\log p  \tnote{***} \atop +1 $ & $2(n+1)\log p + 3n  \atop  +1$ & 4 &  $(4n+3)\log p+3n \atop +2$  \\
\hline \hline
DReLU(n) & $(2n-1)\log p  \tnote{**} \atop +1$ & $ 2n \log p + \atop 3n  -2$  & 4 & $(4n-1) \log p + \atop  3n   -1 $\\
\hline \hline
 SSS(N) for odd N &   $ \log N$       &     $ 2(1+\log N)$         &  1 & $2+3 \log N$    \\
  SSS(N) for even N &   $\log N+1$       &     $ 2(2+\log N)$         &  1 & $ 5+3 \log N$    \\
 SS(N) for odd N &   $ \log N$       &     $ 2(1+\log N)$         &  1 & $2+3 \log N$    \\
  SS(N) for even N &   $\log N+1$       &     $ 2(2+\log N)$         &  1 & $ 5+3 \log N$    \\
\hline \hline
 ReLU(n) &   $(2n-1)\log p  \tnote{**} \atop +n+2$       &     $2n \log p + \atop 5n  + 2$     &  5 & $(4n-1) \log p + \atop  6n +4$    \\
 \hline \hline
\end{tabular}
\begin{tablenotes}
        \footnotesize
        \item[**]  $p\geq n+2$ is a prime number. 
        \item[***]  $p\geq n+3$ is a prime number. 
      \end{tablenotes}
    \end{threeparttable}
\end{center}
\end{table*}

\section{MPC protocols for G-module}
\label{S_G_module}

In this section, we give three MPC protocols for G-module: the G-module action protocol, the shared G-module action protocol and the G-module recover protocol. The G-module action protocol is not used in following part of this paper directly, but it helps us to understand the shared G-module action protocol and the G-module recover protocol.

\subsection{The MPC protocol for the G-module action}
Let $G$ be a finite group, and $A$ be a finite $G$-module. In this subsection, we introduce
Algorithm \ref{alg:G-Module} which describes our 3-party protocol for securely  realizing the functionality $\mathcal{F}_{GM}$ that computes 
$ga\in A$.
At the start of the protocol, parties $P_0, P_1$ hold $g \in G$, $a \in A$ respectively. At the end of protocol, parties $P_0, P_1$ will get shares of $ga$ over $A$.  

The idea is $ga=g(a-b)+gh^{-1}hb=g(a-b)+gh^{-1}u_0+gh^{-1}u_1$, for $g, h\in G$ and $a, b, u_0, u_1 \in A$ satisfying $hb=u_0+u_1$.

\begin{algorithm}[htbp]
  \caption{\textbf{G-Module Action}: \textbf{GM(G,A)}, $\prod _{GM} (\{P_0, P_1\}, P_2)$ }
  \label{alg:G-Module}
  \begin{algorithmic}[1]
    \Require
      $P_0$ holds an element $g\in G$ and $P_1$ holds an element $a \in A$.
    \Ensure
      $P_0$, $P_1$ obtain the shares of $ga$ over $A$.
 \State $P_2$ generates random $h \in G$ and $b \in A$, and splits $hb$ as  $u_0+u_1\in A$;
 \State $P_2$ sends $h$ and $u_0$ to $P_0$ while sends $b$ and $u_1$ to $P_1$;
 \State $P_0$ computes $f:=gh^{-1} \in G$ and then sends it to $P_1$; 
 \State $P_1$ computes $c:=a-b \in A$ and then sends it to $P_0$; 
    \State $P_0$ computes $v_0=gc+fu_0 \in A$ 
        \State $P_1$ computes $v_1=fu_1 \in A$  \\
    \Return $(v_0, v_1)$.
  \end{algorithmic}
\end{algorithm}

This protocol needs $2$ rounds, and its communication is $2(\log |G|+2 \log |A|)$ bits.

But if we use the PRF improvement, we need $\log |A|$ bits communication in offline phase, and $\log |G|+ \log |A|$ bits communication in $1$ round in online  phase per calling.

In fact, $P_0$, $P_1$ can get $(h,u_0)$ and $(b, u_1)$ as follows: Let$P_0$, $P_1$ and $P_2$ share a PRF
\[
\begin{array}{rclcl}
F: K & \times  & \mathbb{Z}/N \mathbb{Z} & \longrightarrow  &G \times A \\
(k&, &i) & \longmapsto & F_k(i)
\end{array}
\]
 Let $P_0$ and $P_2$ share a key $k_0$, $P_0$ and $P_1$ share a key $k_1$. In offline phase, for $i\in \mathbb{Z}/N \mathbb{Z}$, $P_2$ uses $k_0$ as a PRF key to generate $(h_i, u_{i,0})=F_{k_0}(i)$ and uses $k_1$ as a PRF key to generate $\tilde{h}_i, b_i)=F_{k_1}(i)$, then computes $u_{i.1}:=h_ib_i-u_{i,0}$ and finally sends it to $P_1$. $P_1$ stores $ \{u_{i,1} \} _i$.
  In $i$th calling this protocol in online phase, $P_0$ uses $k_0$ to generate  $(h_i, u_{i,0})=F_{k_0}(i)$, and $P_1$ uses $k_1$ to generate $(\tilde{h}_i, b_i)=F_{k_1}(i)$, and restores $u_{i,1}$.
  Hence, the offline communication of GM(G,A) protocol is $\log |A|$ bits per calling (for sending $u_{i,1}$ to $P_1$), the online communication is $\log |G|+ \log |A|$ bits in $1$ round per calling.

The communication and round of GM(G,A) is in Table  \ref{T_all_protocols}.


\subsection{The MPC protocol for the shared G-module action}

Let $G$ be a finite abelian group, and $A$ be a finite $G$-module. In this subsection, we introduce
Algorithm \ref{alg:CG-Module} which describes our 3-party protocol securely realizing the functionality $\mathcal{F}_{SGM}$ that computes 
the result of a pair $(a_0, a_1) \in A^2$  under the shared G-module action.
At the start of the protocol, $P_0$ holds $g_0 \in G, a_0 \in A$, $P_1$ holds $g_1 \in G, a_1 \in A$. At the end of the protocol, parties $P_0$, $P_1$ get
the shares of $g_0g_1(a_0+a_1)$ over $A$.

The principle is given as
\[
\begin{array}{ll}
& g_0g_1(a_0+a_1) \\
= & g_0g_1(a_0+a_1-b_0-b_1)+g_0g_1(b_0+b_1) \\
= & g_0 g_1 (a_1-b_1)+ g_1 g_0 (a_0-b_0)  \\ 
& + g_0g_1 h_0 ^{-1}h_1^{-1} h_0h_1(b_0+b_1) \\
= &  g_0 g_1 (a_1-b_1)+ g_1 g_0 (a_0-b_0)  \\
&+ g_0 h_0 ^{-1} g_1 h_1 ^{-1} h_0h_1(b_0+b_1)
\end{array}
\]
where $g_0, g_1, h_0, h_1 \in G$ and $a_0, a_1, b_0, b_1 \in A$.

\begin{algorithm}[htb]
  \caption{\textbf{shared G-Module Action}: \textbf{SGM(G,A)}, $\prod _{SGM} (\{P_0, P_1\}, P_2)$  }
  \label{alg:CG-Module}
  \begin{algorithmic}[1]
    \Require
      $P_0$ holds  $g_0 \in G$, $a_0 \in A$, $P_1$ holds $g_1 \in G, a_1 \in A$.
    \Ensure
      $P_0$, $P_1$ get the shares of $g_0a_1+g_1a_0$ over $A$.
 \State $P_2$ generates random $h_0, h_1 \in G$ and $b_0, b_1\in A$, and splits $h_0h_1(b_0+b_1)$ as  $u_0+u_1\in A$;
 \State $P_2$ sends $h_0, b_0$ and $u_0$ to $P_0$ and sends $h_1, b_1$ and $u_1$ to $P_1$, respectively;
 \State $P_0$ computes $f_0:=g_0 h_0^{-1} \in G, c_0:=g_0(a_0-b_0) \in A $; 
 \State $P_1$ computes $f_1:= g_1 h_1 ^{-1} \in G, c_1:=g_1(a_1-b_1) \in A$;
 \State $P_0, P_1$ exchange $f_0, f_1, c_0, c_1$, and compute $f:=f_0f_1 \in G$
    \State $P_0$ computes $w_0:=g_0 c_1 +f u_0 \in A$, $P_1$ computes  $w_1:=g_1 c_0 + f u_1 \in A$ \\
    \Return $(w_0, w_1)$.
  \end{algorithmic}
\end{algorithm}

This protocol needs $2$ rounds, and its communication is $ 4\log |G|+6\log |A| $ bits.
But if we use the PRF improvement, similarly to the protocol GM(G,A), we shall need $\log |A|$ bits communication in offline phase, and  $2\log |G|+ 2 \log |A|$ bits  communication in $1$ round in online phase per calling. Hence the communication can be presented as the  Table  \ref{T_all_protocols}.


\subsection{The MPC protocol for the G-module recover}
Let $G$ be a finite group, and let $A$ be a finite $G$- module. Under the action of $G$, $A$ has the $G$-orbit decomposition \cite{Group_action}, \cite{Group_orbit_decomp} as below:
\[
A=\coprod _i A_i
\]
where $A_i=G a_i$ can be generated by any single element $a_i \in A_i$ under the Group action of $G$.

Algorithm \ref{alg:G-Module_R} describes our 3-party protocol for securely realizing the functionality $\mathcal{F}_{GMR}$ that compute
the G-module recover.
At the start of the protocol,
$P_0, P_1$ hold shares $(b_0, b_1)$ of an element $b \in A$ over $A$, and have a common information on the orbit of $b$ under the $G$-action. 
At the end of the protocol, $P_0$ will get an element $g\in G$ and $P_1$ will get an element  $a \in Gb \subset A $ (here $Gb$ is the orbit of $b$ under the $G$-action) such that $ga=b$.

The idea comes from the following equation
\[
g^{-1}(b_0+b_1)=g^{-1}(b_0+b_1-v_0-v_1)+g^{-1}(v_0+v_1)
\]
and the algorithm is given as below:
\begin{algorithm}[htb]
  \caption{\textbf{G-Module Recover}: \textbf{GMR(G,A)}, $\prod _{GMR} (\{P_0, P_1\}, P_2)$  }
  \label{alg:G-Module_R}
  \begin{algorithmic}[1]
    \Require
      $P_0, P_1$ hold shares $(b_0, b_1)$ of an element $b \in A$ over $A$, and a common information on the orbit $B:=Gb$ of $b$ under the group action of $G$.
    \Ensure
      $P_0$ gets an element $g\in G$ and $P_1$ gets an element $a \in Gb $ such that $ga=b$.
 \State $P_2$ generates random $g \in G$ and $u \in A$, and splits $gu$ as $v_0+v_1 \in A$;
 \State $P_2$ sends $g$ and $v_0$ to $P_0$ and sends $u$ and $v_1$ to $P_1$, respectively;
 \State $P_1$ computes $c_1=b_1-v_1 \in A$ and sends it to $P_0$;
    \State $P_0$ computes $w=g^{-1}(b_0-v_0+c_1)\in A$ and sends it to $P_1$;
    \State $P_1$ computes $a:=w+u \in A$;\\
    \Return $(g, a)$.
  \end{algorithmic}
\end{algorithm}

{\noindent\bf Proof of the security of the Algorithm 3:}
It is easy to see that  the conditional distribution $P(c_1=x| g, v_0, b_0, B, b_1)=P(v_1=b_1-x | g, v_0, b_0, B)=P(v_1=b_1-x | g, v_0)$ is a uniform distribution on $A$.
 Hence the posterior distribution $P(b_1 | g, v_0, b_0, B, c_1)$ is equal to the prior distribution $P(b_1 | g, v_0, b_0, B)$.
Therefore $P_0$ can't get any information on $b_1$ from  $(g, v_0, b_0, B, c_1)$.
Similarly since  the conditional distribution $P(w=x| u, v_1, b_1, B, b_0)=P(g^{-1}(b_0+b_1)-u=x| u, v_1, b_1, B, b_0)=P(g^{-1}(b_0+b_1)-u=x | u, B)$ is a uniform distribution on $-u+B$, the posterior distribution $P(b_0 | u, v_1, b_1, B, w)$ is equal to the prior distribution  $P(b_0| u, v_1, b_1, B)$. And
therefore $P_1$ can't get any information on $b_0$ from  $(u, v_1, b_1, B, w)$, which finishes the proof.
\qed

{\noindent\bf Analysis of communication:} Now we give an analysis of the communication of the $G$-module recover.
This protocol needs $2$ rounds, and its communication is $ \log|G|+5\log |A|$ bits.
However if we use the PRF improvement, similarly to the case of GM(G,A), we need only $\log |A|$ bits communication  in offline phase, and $2\log |A|$ bits communication in $1$ round in online  phase per calling which can be shown  in Table 
 \ref{T_all_protocols}.


%

\section{MPC protocol for Assistant OT}
\label{S_AOT}

In this section, we give a 3-party protocol securely realizing the functionality  Oblivious Transfer $\mathcal{F}_{OT}$ for abelian groups. 

Let $A$, $B$ be two finite abelian groups, and let $Map(A, B)$ be the set consisting of all the map from A to B.
There is a natural abelian group structure on $Map(A, B)$ induced from $B$. At the start of the protocol, $P_0$ hold $g \in Map(A, B ) $, $P_1$ hold $j \in A$. At the end of the protocol,
$P_0, P_1$ will get the shares of $g(j)$ over $B$.

For any $k \in A$, let
$$L_k: Map(A, B) \longrightarrow Map(A,B)$$
be the $\lq\lq$left shift" on Map(A, B), which is defined by $L_k(f) (i)= f(i+k)$ with $f\in Map(A, B)$ and $i\in A$. 
The principle of the protocol is
\[
g(i)=(L_{i-\tilde{i}}g-f)(\tilde{i})+f(\tilde{i})
\]
for any $f, g \in Map(A, B)$ and any $i, \tilde{i} \in A$. 
Our protocol is described as below:
\begin{algorithm}[htb]
  \caption{\textbf{Assistant OT}   :  \textbf{AOT(A,B)}, $\prod _{AOT} (\{P_0, P_1\}, P_2)$ }
  \label{alg:Assistant_OT}
  \begin{algorithmic}[1]
    \Require
      $P_0$ holds $i \in A$, $P_1$ holds $g \in Map( A, B ) $
          \Ensure
      $P_0$, $P_1$ gets the share of $g(i) \in B$.
          \State $P_2$ generates  random $\tilde{i} \in A$ and random $f \in Map(A, B)$, and split $f(i)$ to $f(i)=a_0+a_1 \in B$.
          Then  $P_2$ sends $\tilde{i}, a_0$ to $P_0$, and sends $f$, $a_1$ to $P_1$;
    \State $P_0$ computes $k:=i-\tilde{i}$ and sends $k$ to $P_1$;
    \State $P_1$ computes $h:=L_k g - f \in Map(A,B)$, and sends $h$ to $P_1$;
    \State $P_1$ computes $x_0:=h(\tilde{i})+a_0$ \\
    \Return $(x_0, a_1)$.
  \end{algorithmic}
\end{algorithm}

This MPC protocol needs $2$ rounds, and its communication is $ 2|A|\log |B|+ 2 \log |A| + \log |B|$ bits.
But if we use the PRF improvement, similarly to the protocol of GM(G,A), we need only $\log |B|$ bits communication in offline phase, and $|A|\log |B|+ \log |A|$ bits communication in $1$ round in online  phase per calling 
which is shown in  Table  \ref{T_all_protocols}.

\section{MPC protocol for Module Transform}
\label{S_MoT}

In this section, we give a review of a  3-party protocol securely
 realizing the functionality $\mathcal{F}_{MoT}$ for module transform in \cite{similar}.
 
 At the start of the protocol, $P_0, P_1$ hold shares of $a \in \mathbb{F}_2$ over $\mathbb{F}_2$, and a common integral number $m$.  At the end of the protocol, $P_0$, $P_1$ get the shares of $a$  over $\mathbb{Z}/m\mathbb{Z}$.

Let
$$I : \mathbb{F}_2 \longrightarrow \mathbb{Z}/m\mathbb{Z}$$
be the module transform map defined by $I(0)=0, I(1)=1$. 
One can use the protocol to compute $I(a)$. The explicit protocol is as below Algorithm \ref{alg:MoT(m)}.
\begin{algorithm}[htb]
  \caption{\textbf{Module Transform}: \textbf{MoT(m)}, $\prod _{MoT} (\{P_0, P_1\}, P_2)$  }
  \label{alg:MoT(m)}
  \begin{algorithmic}[1]
    \Require
      $P_0, P_1$ hold shares of $a$ over $\mathbb{F}_2$,  and a common integral number $m$.
    \Ensure
      $P_0$, $P_1$ get the shares of $a$ over $\mathbb{Z}/m\mathbb{Z}$.
 \State $P_2$ generates random $u \in \mathbb{F}_2$, then splits $u$ into $(u_0, u_1) \in \mathbb{F}_2 \times \mathbb{F}_2$ and $I (u) \in  \mathbb{Z}/m\mathbb{Z}$ into $(b_0, b_1) \in \mathbb{Z}/m\mathbb{Z} \times  \mathbb{Z}/m\mathbb{Z}$, and finally sends $(u_0, b_0)$ to $P_0$, send $(u_1, b_1)$ to $P_1$ respectively;
 \State $P_0$ computes $z_0:=a_0-u_0$ locally and $P_1$ computes $z_1:=a_1-u_1$ locally;
  \State $P_0$ and $P_1$ reconstruct $z:=a-u$ by interchanging  $z_0$ and $z_1$;
 \State $P_0$ computes $y_0:=(-1)^zb_0 \in \mathbb{Z}/m\mathbb{Z}$ and $P_1$ computes $y_1:=(-1)^zb_1+z \in \mathbb{Z}/m\mathbb{Z}$; \\
    \Return $(y_0, y_1)$.
  \end{algorithmic}
\end{algorithm}

This protocol needs  $2$ rounds, and its communication is $ 2(\log m+2)$ bits.
However if we use the PRF improvement, we need $\log m$ bits communication in offline phase, and  $2$ bits communication in $1$ round in online  phase per calling.

In fact, $P_0$, $P_1$ can obtain $(u_0, b_0)$ and $(u_1, b_1)$ as follows. Let $P_0$, $P_1$ and $P_2$ have a common  PRF improvement by the map
$$
\begin{array}{rclcl}
F: K & \times  & \mathbb{Z}/N \mathbb{Z} & \longrightarrow  & \mathbb{F}_2 \times Map(\mathbb{F}_2,\mathbb{Z}/m\mathbb{Z})\\
(k&, &i) & \longmapsto & F_k(i)
\end{array}
$$
Let $P_0$ and $P_2$ share a key $k_0$, $P_1$ and
$P_2$ share a key $k_1$. In offline phase, for $i\in\mathbb{Z}/N \mathbb{Z}$, $P_2$ shall use $k_0$ as PRF key to generate $(u_{i,0}, b_{i,0})=F_{k_0}(i)$ and use $k_1$ as PRF key to generate $(u_{i,1}, \tilde{b}_{i,1})=F_{k_1}(i)$,
then computes $b_{i.1}:=I(x-u_{i,0}-u_{i,1})-f_{i,0}$, and finally sends $b_{i.1}$ to $P_1$. $P_1$ will store $ \{b_{i.1} \} _{i\in\mathbb{Z}/N \mathbb{Z}}$.
In the $i$-th calling of this protocol in online phase, $P_0$ uses $k_0$ to generate  $(u_{i,0}, b_{i,0})=F_{k_0}(i)$; $P_1$ uses $k_1$ to generate $(u_{i,1}, \tilde{b}_{i,1})=F_{k_1}(i)$, and restore $b_{i,1}$.
Hence, the offline communication of \textbf{MoT}(m) protocol is $\log m$ bits per calling (for sending $b_{i,1}$ to $P_1$),
 and the online communication of \textbf{MoT}(m) protocol is $2$ bits per calling (for interchanging  $z_0$ and $z_1$) as shown in 
 Table  \ref{T_all_protocols}.

%

\section{MPC protocol for security comparison}
\label{S_SC}

%
%
%
%
%
In this section we will supply a 3-party protocol securely realizing the functionality $\mathcal{F}_{SC}$ for security comparison that is as follows:  At the start of the protocol, $P_0, P_1$ hold $x \in \mathbb{Z}/2^n\mathbb{Z}, y \in \mathbb{Z}/2^n\mathbb{Z}$ respectively. At the end of the protocol, $P_0, P_1$ get shares of ($x < y$) over $\mathbb{Z}/2\mathbb{Z}$.

 For that purpose we will first give a 3-party protocol securely realizing the functionality $\mathcal{F}_{FNZ}$ for searching the first non-zero bit in the $\mathcal{F}_{GMR}$-hybrid model.
And then we shall give a 3-party protocol securely realizing the functionality $\mathcal{F}_{SC}$ in the $(\mathcal{F}_{FNZ}, \mathcal{F}_{AOT}, \mathcal{F}_{MoT})$-hybrid model.

For a positive integer $n$, below we shall often use the notation 
$$I_{n}=\{0,1,\cdots,n-1\}$$

\subsection{MPC protocol to search first non-zero bit}

Let $p\geq n+2$ be a prime number, and let $P_0$, $P_1$ hold shares of a non-zero $0$-$1$ vector $u=(u_i)_{i=0}^{n-1}$  over $\mathbb{F}_p^n$. Here $0$-$1$ vector $u=(u_i)_{i=0}^{n-1}$ is a vector satisfying
$$u_i =0 \mbox { or } 1, \forall\; i\in I_n\quad\quad \&\quad\quad\exists\; u_j=1,\;j\in I_n$$
We will give a 3-party protocol to realize the functionality $\mathcal{F}_{FNZ}$ that search the first non-zero bit of $u$. At the end of the protocol,
$P_0$, $P_1$ will get the shares of $\min \{i\in I_n : u_i \neq 0 \}$ over $\mathbb{Z}/n\mathbb{Z}$.

%

Let $G:=\mathbb{Z}/n\mathbb{Z} \ltimes
(\mathbb{F}_p ^\times )^n$ be the semi-direct product of the groups $\mathbb{Z}/{n\mathbb{Z}}$ and $(\mathbb{F}_p^ \times)^n$ (\cite{semidirect_product}, p.195 in \cite{GTM-4}).
The underlying set of the group $G$ is the Cartesian product $\mathbb{Z}/n\mathbb{Z} \times (\mathbb{F}_p ^\times)^n $ while the group operation is defined by

\begin{align*}
G \times  G & \longrightarrow  G \\
\left((i, a),(j, b)\right) & \longmapsto \left(i+j, T_a L_i(b)\right)
\end{align*}
Here $L_i$ is the $i$-th circular left shift operator on $\mathbb{F}_p^n$ and $T_a$ is the $\lq\lq$ multiply by $a$ " operator on $\mathbb{F}_p ^ n$, i.e.,
 for $x=(x_0, x_1, \cdots x_{n-1}) \in \mathbb{F}_p ^ n$,
we have
\[
L_i(x)=(x_i, x_{i+1}, \cdots, x_{n-1}, x_0, \cdots x_{i-1}) \in \mathbb{F}_p ^ n
\]
and
\[
T_a(x)=(a_0x_0, a_1x_1, \cdots, a_{n-1}x_{n-1}) \in \mathbb{F}_p ^ n
\]
respectively.

It is not difficult to verify that $G $ is a non commutative group with the identity element $(0, 1^n )$. One can define the $G-$module structure on $\mathbb{F}_p  ^n$ as follows:
\begin{align*}
G \times\mathbb{F}_p  ^n & \longrightarrow \mathbb{F}_p  ^n \\
((i,a), x)&\longmapsto T_aL_i(x)
\end{align*}
Then we have the following Lemma.

\begin{lem}
\label{orbit_decom}
Let $G:=\mathbb{Z}/n\mathbb{Z} \ltimes
(\mathbb{F}_p ^\times )^n$ be the semi-direct product of the group $\mathbb{Z}/n\mathbb{Z}$ and the group $(\mathbb{F}_p^ \times)^n$. There is a $G$-orbit decomposition 
$$\mathbb{F}_p  ^n=\coprod _{d=0} ^n U_d$$
of $\mathbb{F}_p  ^n$, where
$U_d$ is the subset of  $\mathbb{F}_p  ^n$ consisting of the elements of Hamming weight $d$.
\end{lem}

Now we will give our 3-party protocol to realizing the functionality $\mathcal{F}_{FNZ}$ that computing the first non-zero bit of $u$ securely in the $\mathcal{F}_{GMR}$-hybrid model. The main idea comes  from the following lemma:

\begin{lem}
\label{only_zero}
Let $p\geq n+2$ be a prime number, and let $u=(u_i)_{i=0}^{n-1}$ be a non-zero $0$-$1$ element in $\mathbb{F}_p^n$. 
Let $v \in \mathbb{F}_p ^n$ defined as
\[
\begin{array}{l}
v_0=u_0  \\
v_i=v_{i-1}+u_i \quad \mbox{ for } i = 1, 2, \cdots, n-1
\end{array}
\]
Thus $v_i \in I_{n+1}$ for all $i\in I_n$.
Let $f$ be a map
\begin{align*}
f :  \{0,1\} \times I_{n+1} & \longrightarrow \mathbb{F}_p  \\
(a, b) & \longmapsto  b-2a+1\mod p
\end{align*}
Then we have $\min \{ i| u_i \neq 0, i\in I_n\}$ is the unique $i \in I_n$ such that $f(u_i, v_i)=0\mod p$.
\end{lem}
{\noindent\bf Proof.}
First we claim that $(1,1)$ is the unique $(a,b)\in\{0,1\} \times I_{n+1}$ such that $f(a,b)=0\mod p$.
That is because if $a=0$, $f(a,b) \in \{1, 2, \cdots n+1\}$ for $b \in I_{n+1}$,  which implies $f(a,b)\neq0\mod p$; while if $a=1$, then $b=1$ is the only solution such that of $f(1,b)=0\mod p$.

Now it is not difficult to see that  $\min \{ i=0, 1, \cdots, n-1 | u_i \neq 0\}$ is the unique $i \in I_n$ such that both $u_i=1$ and $v_i=1$. Thus $\min \{ i| u_i \neq 0,\; i\in I_n\}$ is the unique 
$i\in I_n$ such that $f(u_i, v_i)=0$ which finishes  the proof.
\qed


\vskip15pt

Following Lemma \ref{only_zero}, we design a 3-party protocol to compute the first non-zero bit of a non-zero $0$-$1$ vector $u=(u_i)_{i=0}^{n-1}$ in $\mathbb{F}_p^n$, where the input is its shares over $\mathbb{F}_p^n$, and the output is  its shares 
over $\mathbb{Z}/n\mathbb{Z}$. The principle is not difficult: Let $v$ and $f$ as in Lemma \ref{only_zero},  $z=\left(f(u_0,v_0)), \cdots, f(u_{n-1}, v_{n-1})\right)\in\mathbb{F}_p^n$, and the group $G$ as in
Lemma \ref{orbit_decom}, then  the orbit  $Gz$
will be the unique $G$-orbit of Hamming weight $n-1$ in the
decomposition in Lemma \ref{orbit_decom}, which is a common information for each parts. If there is a $g=(i, c) \in G$ and $w \in \mathbb{F}_p^n$ such that $gw=z$, then the first non-zero bit of $z$ is $(i+j) \mod n$, where $j$ is the first non-zero bit of $w$.

Now we give our 3-party protocol to realize the functionality $\mathcal{F}_{FNZ}$ securely in $\mathcal{F}_{GMR}$-hybrid model 
 in Algorithm \ref{alg:FNZ}:

\begin{algorithm}[htb]
  \caption{\textbf{First non-zero bit}: \textbf{FNZ(p, n)}, $\prod _{FNZ} (\{P_0, P_1\}, P_2)$ }
  \label{alg:FNZ}
  \begin{algorithmic}[1]
    \Require
     Let $p\geq n+2$ be a prime number, $P_0, P_1$ hold shares of a non-zero $0$-$1$ vector $u=(u_i)_{i=0}^{n-1}\in\mathbb{F}_p^n$
    \Ensure
      $P_0$, $P_1$ obtain the shares of $\min \{i| u_i \neq 0, i\in I_n\}$ over the abelian group $\mathbb{Z}/n\mathbb{Z}$.
    \State $P_0, P_1$ compute $v_i:= \sum _{j=0} ^i u_i  \in \mathbb{Z}/p\mathbb{Z}$ locally for $i\in I_n$;
    \State $P_0, P_1$ compute the shares of $z_i=f(u_i,v_i)$ over $\mathbb{Z}/p\mathbb{Z}$ locally for $i\in I_n$;
	\State  $P_0, P_1$ run the GMR($G$, $\mathbb{F}_p^n$) protocol, and $P_0$ get an elements $g=(i, c) \in G$, $P_1$ obtain an element $w$ such that $gw=z$;
	\State $P_1$ take the only $j \in I_n$  such that $w_j=0$\\
    \Return $(i, j)$.
  \end{algorithmic}
\end{algorithm}

The round and communication of the protocol FNZ(p, n) are the same as those of GMR(G, $\mathbb{F}_p^n$), where $G=\mathbb{Z}/n\mathbb{Z} \ltimes
(\mathbb{F}_p ^\times )^n$. Hence its offline communication is $n\log p$, its online communication is $2n\log p$ in $1$ round.
Its communication and round is shown in  in Table  \ref{T_all_protocols}.
%
%

\subsection{MPC protocol for security comparison}

In this subsection we will give a 3-party protocol realizing the functionality 
security comparison $\mathcal{F}_{SC}$ securely in $(\mathcal{F}_{FNZ}, \mathcal{F}_{MoT}, \mathcal{F}_{AOT})$-hybrid model . The idea is that, for two non-zero $0$-$1$ elements $x=(x_0,x_1,\cdots,x_{n-1})$ and $y=(y_0,y_1,\cdots,y_{n-1})$,
if $i$ is the rightmost bit such that $x_i\neq y_i$, then $(x<y)==(y_i=1)$.

The algorithm is as below Algorithm \ref{alg:ED}:
\begin{algorithm}[htb]
  \caption{\textbf{Secure compare}: \textbf{SC(n)}, $\prod _{SC} (\{P_0, P_1\}, P_2)$ }
  \label{alg:ED}
  \begin{algorithmic}[1]
    \Require
      $P_0$ holds $x\in\mathbb{Z}/ 2^n\mathbb{Z}$, $P_1$ holds $y\in\mathbb{Z}/ 2^n \mathbb{Z}$. $P_0, P_1$ have the common
      information that $p\geq n+3$ is a prime number.
    \Ensure
      $P_0$, $P_1$ get the shares of ($x < y$) over $\mathbb{F}_2$
      \State $P_0$ writes $x$ as the binary representation $(x_i)_{i=0}^{n-1}$ such that $x=\sum\limits_ {i=0} ^{n-1} x_i 2^{n-1-i}$, 
      $P_1$ writes $y$  as the binary representation $(y_i)_{i=0} ^{n-1}$ such that $y=\sum\limits_{i=0} ^{n-1} y_i2^{n-1-i}$;
      \State $P_0$ puts $x_n=1$, $P_1$ puts $y_n=0$;
          \State  For each bit $i\in I_n$, $P_0$, $P_1$ and $P_2$ run the \textbf{MoT}(p) protocol for $(x_i, y_i)_{i=0} ^{n-1}$, which is the shares of $(x_i+y_i \mod 2) _{i=0} ^{n-1} = (x_i \neq y_i) _{i=0} ^{n-1}$. Then $P_0$, $P_1$ get the shares of $u_i=(x_i \neq y_i)$ over $\mathbb{F}_p$ for $i\in I_n$;
          \State Let $P_0, P_1$ hold the first and second component of the shares $(1,0)$ of $u_n=1$ separately;
    \State $P_0$ and $P_1$ run the  \textbf{FNZ(p, n+1)} protocol for $(u_i)_{i=0}^n$ and obtain the shares $(i_0, i_1)$ of  $\min \{i| u_i=0, i\in I_{n+1}\}  \in \mathbb{Z}/(n+1)\mathbb{Z}$;
    \State $P_1$  computes  the  circular left shift $L_{i_1}  (y_i)_{i=0}^{n} $;
    \State $P_0$ and $P_1$ run the protocol \textbf{AOT}($\mathbb{Z}/(n+1)\mathbb{Z}$, $\mathbb{F}_2$), and then obtain the shares $(z_0, z_1)$ of $(L_{i_1} (y_i)_{i=0}^{n} )_{i_0}=y_{i_0+i_1 \mod (n+1)}$ over $\mathbb{Z}/2\mathbb{Z}$\\
    \Return $(z_0, z_1)$.
  \end{algorithmic}
\end{algorithm}

{\noindent\bf Communication analysis:}
The \textbf{SC(n)} protocol uses $n$ \textbf{MoT}($p$) in 1 round, 1 \textbf{FNZ($p, n+1$)} in 2 round and $1$ \textbf{AOT}($\mathbb{Z}/(n+1)\mathbb{Z}$, $\mathbb{F}_2$) in 2 round, where $p$ is a prime number with $p\geq n+2$. Hence its
communication equal to $n \times \mathbf{MoT}(o) + \mathbf{FNZ}(p, n+1) + \mathbf{AOT}(\mathbb{Z}/(n+1)\mathbb{Z}, \mathbb{F}_2)$ and the round complex is 5.
. 
But we can reduce the communication more. In fact, the $i_0$ in Step 5 is chosen by $P_2$. Hence, $P_2$
can use this $i_0$ as $\tilde{i}$ in the offline part of Step 5 still. Then the massage $k:=i-\tilde{i}$ in the Step 2 in the protocol AOT(A, B) is always equal to $0$ and is need not be sent. Hence the communication is reduced to $n \times \mathbf{MoT}(o) + \mathbf{FNZ}(p, n+1) + \mathbf{AOT}(\mathbb{Z}/(n+1)\mathbb{Z}, \mathbb{F}_2) - \log (n+1)$, the round complex is reduce to $4$.

We shown the communication of our \textbf{SC(n)} in the Table \ref{T_all_protocols}.
%
%

Table \ref{T_Compare_SC} gives a comparison between our protocol and some known protocols, for example, those in \cite{FSS,similar,GSV07,KSS09}. Obviously our communication is much less.

\begin{table*}[htbp]
\caption{Compare to exists  SC protocols}
\begin{center}
\begin{threeparttable}
\begin{tabular}{c|c|c|c|c|c}
\hline
 n & Protocol  & offline com.  & online comm.  & online round & total comm.\\ \hline
 n & \color{red}{Our}    & $\color{red}{(2n+1)\log p  +1} \tnote{*}  $  & \color{red}{$2(n+1) \log p + \atop 3n + 1$}  & \color{red}{3} & \color{red}{$(4n+3) \log p + \atop  3n  +2$}\\
n & FSS \cite{FSS} &  $\approx 2\lambda n$  \tnote{**} &  2n & 1 &  $\approx 2\lambda n + 2n$ \\
n & NPSETC SC1\cite{similar} & $O(kn/\log k) \mbox{ if } n = o(k^2)  \atop O(n) \mbox{ else }$ & $O(n)$ &  $O(\log \log n)$ &  $O(n)$  \\
n & NPSETC SC2\cite{similar} & $O(kn/\log k) \mbox{ if } n^{1-1/c} = o(k^2)  \atop O(n) \mbox{ else }$ & $O(n)$ &  $O(c \log ^* n)$ &  $O(n)$  \\
n & NPSETC SC3\cite{similar} & $O(kn/\log k) \mbox{ if } n^{1-1/c} = o(k^2)  \atop O(n) \mbox{ else }$ & $O(n)$ &  $O(c \log ^* n)$ &  $O(n)$  \\
\hline
 32 &\color{red}{ Our}    & \color{red}{340}  & \color{red}{441} & \color{red}{3} & \color{red}{781} \\
32 & FSS \cite{FSS} &  $\approx  4096 \times 2$   &  64 & 1 &  $\approx 8256$  \\
32 & NPSETC SC1\cite{similar} & 15120 &  530 &  12 & 15650\\
32 & NPSETC SC2\cite{similar} & 12568 &  3125 & 7 & 15693 \\
32 & NPSETC SC3\cite{similar} & 12394 & 622 & 10 & 13016 \\
32 & GSV07  \cite{GSV07} & 14062 &  1068 & 6 & 15130 \\
32 & KSS09 \cite{KSS09} & 12352 & 12320 & 2 & 24672 \\
\hline
 64 & \color{red}{Our}    & \color{red}{784}  & \color{red}{982} & \color{red}{3} & \color{red}{1766}\\
64 & FSS \cite{FSS} &  $ 8512 \times 2$   &  128 & 1 &  $ 17152$  \\
64 & NPSETC SC1\cite{similar} & 31388 & 1120 &  12 & 32508\\
64 & NPSETC SC2\cite{similar} & 28872 &  4138 & 7 & 33010 \\
64 & NPSETC SC3\cite{similar} & 28786 & 1286 & 10 & 30072 \\
64 & GSV07  \cite{GSV07} & 29072 & 2208 & 7 & 31280 \\
64 & KSS09  \cite{KSS09} & 24804 & 24640 & 2 & 49344 \\
\hline
 128 & \color{red}{Our}    & \color{red}{1809}  & \color{red}{2200} & \color{red}{3} & \color{red}{4009}\\
128 & FSS \cite{FSS} &  $ \approx 16384 \times 2$   &  256 & 1 &  $ \approx 33024$  \\
128 & NPSETC SC1\cite{similar} & 52121 & 2101 &  12 &  54222\\
128 & NPSETC SC2\cite{similar} & 48031 &  5801 & 7 &  53832\\
128 & NPSETC SC3\cite{similar} & 47963 & 2239 & 10 & 50202 \\
128 & GSV07 \cite{GSV07} & 59250 & 4500 & 8 & 63750\\
128 & KSS09 \cite{KSS09} & 49408 & 49280 & 2 & 98688 \\
\hline
\label{T_Compare_SC}
\end{tabular}

\vskip15pt
\begin{tablenotes}
        \footnotesize
        \item[*]  Here $p$ is a prime number with $p\geq n+3$. \\
        \item{**} In paper \cite{FSS}, $\lambda=128$.
      \end{tablenotes}
    \end{threeparttable}
\end{center}
\end{table*}

%
%
%
%
%
%
%
%
%

\section{MPC protocol for DReLU}
\label{S_DReLU}

In fixed point representation of real number, we usually use two's complement to represent a negative number, hence in order to confirm 
a number $x \in \mathbb{Z}/2^n\mathbb{Z}$ is not $\lq\lq$ negative ", we need to check whether $x<2^{n-1}$ or not.

In the share representation of $x= u+ v\in\mathbb{Z}/ 2^n \mathbb{Z}$, one can write $u$ and $v$ in the binary form
\begin{align*}
&u=u_0+u_1*2+ \cdots u_{n-1}*2^{n-1},\\
&v=v_0+v_1*2 + \cdots v_{n-1}*2^{n-1}
\end{align*}
where $u_i, v_i\in\{0,1\}$ for all $i\in I_n$. In terms of the binary form of $u$ and $v$, we shall use the notation
$\tilde{u} = u_0+u_1*2+ \cdots u_{n-2} * 2^{n-2}$ and $\tilde{v}=v_0+v_1 *2 + \cdots v_{n-2} *2^{n-2}$ respectively.

Now we define $P$, $Q$ be two elements in $\mathbb{Z}/2\mathbb{Z}$ as
\[
\begin{array}{ll}
P:= & (( \tilde{u} + \tilde{v} ) \geq 2^{n-1}  )  \mbox{     (boolean expression)  }\\
Q:= & (u_{n-1}+ v_{n-1}) \mod 2
\end{array}
\]
Then we get the following lemma.

\begin{lem}
\label{SCtoSReLU}
The boolean value of ($x<2^{n-1}$) is equal to $1+P+Q \mod 2$ under the identities true $=1$ and false $=0$.
\end{lem}

{\noindent\bf Proof:} Under the identities true $=1$ and false $=0$, we have
\[
\begin{array}{ll}
 & (x \geq 2^{n-1}) \\
= &u_{n-1}+v_{n-1}+\mbox{ carry of } \tilde{u}+\tilde{v} \\
= & Q+ P \mod 2
\end{array}
\]
Hence we have
\[
(x<2^{n-1}) = 1+P+Q \mod 2.
\]
\qed

Based on the Lemma \ref{SCtoSReLU}, algorithm \ref{alg:DReLU} describes our 3-party protocol securely realizing the functionality $\mathcal{F}_{DReLU}$   in the $\mathcal{F}_{SC}$ -hybrid model.

\begin{algorithm}
  \caption{\textbf{DReLU(n)}, $\prod _{DReLU} (\{P_0, P_1\}, P_2)$ }
  \label{alg:DReLU}
  \begin{algorithmic}[1]
    \Require
      $P_0, P_1$ hold shares $ (u, v)$ of $x$ over $\mathbb{Z}/ 2^n \mathbb{Z}$.
    \Ensure
      $P_0$, $P_1$ get the shares of ($x < 2^{n-1}$) over  $\mathbb{Z}/2\mathbb{Z}$;
 \State $P_0$ has the $0 \sim n-2$ bits of $\tilde{u}$  and the last bit $u_{n-1}$ of $u$, $P_1$ has the $0 \sim n-2$ bits of $\tilde{v}$ and the last bit $v_{n-1}$ of $v$ respectively;
 \State $P_0, P_1$ and $P_2$ call  $\mathbf{SC}(n-1)$ for $(2^{n-1}-1- \tilde{u}, \tilde{v})$ and get the shares of  $P:=(( \tilde{u} + \tilde{v} ) \geq 2^{n-1} )$ over $\mathbb{Z}/2\mathbb{Z}$;
 \State $P_0, P_1$ take $(u_{n-1}, v_{n-1})$ as the shares of $Q:= u_{n-1}+v_{n-1} \mod 2$ over $\mathbb{Z}/2\mathbb{Z}$;
    \State $P_0, P_1$ compute  the shares $w$ of $1+P+Q$ over $\mathbb{Z}/2\mathbb{Z}$;\\
    \Return $w$.
  \end{algorithmic}
\end{algorithm}

\begin{rem}
In \cite{secureNN}, \cite{CrypTFlow}, the protocol $\prod _{DReLU}$ realizes the functionality $\mathcal{F}_{DReLU}$ for $x$ in the
subdomain $\left[0, 2^k\right] \cup \left[2^n-2^k, 2^n -1\right]$ of $\left[ 0, 2^n-1 \right]$, where $k<n-1$. But our protocol  realizes the functionality $\mathcal{F}_{DReLU}$ for all the $x$ in $\left[ 0, 2^n-1 \right]$.
\end{rem}

\begin{rem}
In our protocol, the assistant third part $P_2$ only sends message to the parties $P_0, P_1$ in offline phase. In online phase,
just $P_0$ and $P_1$ play the protocol. But in \cite{secureNN}, \cite{CrypTFlow},  the assistant third part $P_2$ needs to both receive and send message in online  phase.
\end{rem}

The communication of protocol \textbf{DReLU}(n) is same as the protocol  \textbf{SC}(n-1), and is shown in Table \ref{T_all_protocols}.
We compare our protocol \textbf{DReLU}(n) with some exists protocol in Table \ref{T_Compare_DReLU} also. Note that in  \ref{T_Compare_DReLU}, $p$ is a prime number greater than or equal to $n+2$.

\begin{table*}[htbp]
\caption{Comparison to exists  DReLU protocols}
\begin{center}
\begin{threeparttable}
\begin{tabular}{c|c|c|c|c|c}
\hline
 n & Protocol  & offline com.  & online comm.  & online round & total comm.\\ \hline
 n & \color{red}{Our}    & \color{red}{$(2n-1)\log p  +1$} \tnote{*}  & \color{red}{$2n \log p + \atop 3n  -2$}  & \color{red}{3} & \color{red}{$(4n-1) \log p + \atop  3n  -1$}\\
 n & CrypTFlow \cite{CrypTFlow}  & 0 &  8n log p + 14n & 8 & 6n log p + 14n \\
n & SecureNN \cite{secureNN}  & 0 &  8n log p + 19n & 8 & 8n log p + 19n \\
\hline
 32 & \color{red}{Our}    & \color{red}{329.2}  & \color{red}{426.4} & \color{red}{3} & \color{red}{756.6} \\
32 & SecureNN \cite{secureNN}  & 0 &  1448.3 & 8 & 1448.3 \\
32 & SecureNN \cite{secureNN}  & 0 &  1941.6 & 8 & 1941.6 \\
\hline
 64 & \color{red}{Our}    & \color{red}{771.4}  & \color{red}{966.5} & \color{red}{3} & \color{red}{1737.9}\\
64 & CrypTFlow \cite{CrypTFlow} & 0 & 3225.4 & 8 & 3225.4 \\
64 & SecureNN \cite{secureNN}  & 0 &  4321.8 & 8 &  4321.8\\
\hline
 128 & \color{red}{Our}    & \color{red}{1794.5}  & \color{red}{2182.6} & \color{red}{3} & \color{red}{3977.1}\\
 128 & CrypTFlow \cite{CrypTFlow}  & 0 & 7193.7 & 8 & 7193.7 \\
128 & SecureNN \cite{secureNN}  & 0 & 9634.2  & 8 &  9634.2\\

\hline
\label{T_Compare_DReLU}
\end{tabular}
\begin{tablenotes}
        \footnotesize
        \item[*]  $p\geq n+2$ is a prime number. \\
        \item{**} In paper \cite{FSS}, $\lambda=128$.
      \end{tablenotes}
    \end{threeparttable}
\end{center}
\end{table*}

\section{MPC protocol for Select Shares}
\label{S_SS}

Let $x, y \in \mathbb{Z}/N\mathbb{Z}$, $a \in  \mathbb{F}_2$.
We will give our MPC protocol realizing the functionality Select Shares $\mathcal{F}_{SS}$. 
At the start of the protocol, two parties $P_0$  and $P_1$ hold shares of $x$, $y$ over $\mathbb{Z}/N\mathbb{Z}$
and shares of  $a \in \mathbb{F}_2$ over $\mathbb{F}_2$.
 At the end of the protocol, $P_0$ and $P_1$ will learn the shares of $s$  over $\mathbb{Z}/N\mathbb{Z}$
which is defined as
\[
s:=
\left \{
\begin{array}{ll}
x & \mbox{ if } a=1, \\
y & \mbox{ if } a=0.
\end{array}
\right.
\]
Since $s$ depends on $a$, we shall call $a$ the selection bit.

Note that $s=a(x-y)+y$. Hence this functionality can be reduced to the spacial case that $y=0$ without any communication. Hence we only need to realize the
special select share functionality $\mathcal{F}_{SSS}$: At the beginning parties $P_0$, $P_1$ hold shares of both $z \in \mathbb{Z}/N\mathbb{Z}$ over  
$\mathbb{Z}/N\mathbb{Z}$ and a selection bit $a \in \mathbb{F}_2$ over $\mathbb{F}_2$.
At the end of protocol they will get the shares of
$az:=z \mbox{ if } a=1 \mbox{ otherwise }0$ over $\mathbb{Z}/N\mathbb{Z}$.

In \cite{secureNN}, the matrix multiplication protocol is used to realize the select share functionality. However we shall use the G-module action protocol to do this. 
By our protocol, the communication is highly cut down.

The principle of our protocol is mainly based on the equation
\[
az = \frac{z-(-1)^az}{2} \mbox{ for } a=0,1, \mbox{ and } z\in \mathbb{Z}.
\]

\subsection{Special select share protocol for odd module}

Let $a=a_0+a_1 \mod 2$, $z=z_0+z_1 \mod N$ be the shares of $a$ over $\mathbb{Z}/2\mathbb{Z}$ and $z$ over $\mathbb{Z}/N\mathbb{Z}$ respectively, where $N$ is an odd number.
Then we have
\begin{align*}
(-1)^a z &= (-1)^{a_0+a_1}(z_0+z_1)\\
&= (-1)^{a_0} (-1)^{a_1}(z_0+z_1) \in \mathbb{Z}/N\mathbb{Z}
\end{align*}
.

%
%
%

Let $G:=\{ \pm 1\}$, $A:=\mathbb{Z}/N\mathbb{Z}$. It easy to see that $A$ is a $G$-module. Hence we can use the shared G-module action protocol to compute 
$(-1)^az$ and hence to compute $z-(-1)^az \in \mathbb{Z}/N\mathbb{Z}$.
In the case that $N$ is an odd number, then $2$ is invertible in $Z/NZ$, and hence it easy to compute $zx$ from  $z-(-1)^az$.
The algorithm is given as Algorithm \ref{alg:SSS_odd}.

\begin{algorithm}[htb]
  \caption{Special Select Share \textbf{SSS(N)} for an odd number N, $\prod _{SSS} (\{P_0, P_1\}, P_2)$}
  \begin{algorithmic}[1]
    \Require
      $P_0, P_1$ hold the share representation of $z\in \mathbb{Z}/N\mathbb{Z}$ and $a \in \mathbb{F}_2$
    \Ensure
      $P_0$, $P_1$ get the share representation of $az \in \mathbb{Z}/N\mathbb{Z}$.
 \State $P_0, P_1$ run the \textbf{SGM}($\{ \pm 1 \}, \mathbb{Z}/N\mathbb{Z}$) protocol  to get the share representation
 $(u_0, u_1)$ of $(-1)^az \in \mathbb{Z}/N\mathbb{Z}$;
 \State $P_0$ computes $v_0:= \frac{z_0-u_0}{2}  \mod N$, $P_1$ computes $v_1:=  \frac{z_1-u_1}{2}   \mod N$\\
     \Return $(v_0, v_1)$.
  \end{algorithmic}
  \label{alg:SSS_odd}
\end{algorithm}

\subsection{Special Select Shares protocol for even module}

In the case that $N$ is an even number, $2$ is not invertible in $Z/NZ$, we need to modify the protocol.

In fact, if $N$ is an even number, $a=a_0+a_1 \mod 2$, $z=z_0+z_1 \mod N$ are the share representations of $a$ and $z$ respectively,
one can lift $z_0, z_1$ to $\tilde{z}_0, \tilde{z}_1 \in \mathbb{Z}/2N\mathbb{Z}$ respectively. Let
$\tilde{z}:=\tilde{z}_0+\tilde{z}_1 \mod 2N$, then we have $az \equiv a\tilde{z} \mod N$ for $a=0$, $1$.


Using the same method as in the case $N$ is an odd number, we can get the share representations of $2a\tilde{z}=\tilde{z}-(-1)^a\tilde{z} \in \mathbb{Z}/2N\mathbb{Z}$,
and $a \tilde{z} \in \mathbb{Z}/N\mathbb{Z}$ which is equal to $az \in \mathbb{Z}/N\mathbb{Z}$. And the algorithm is shown as Algorithm \ref{alg:SSS_even}.

\begin{algorithm}[htb]
  \caption{Special Select Share \textbf{SSS(N)} for even number N, $\prod _{SSS} (\{P_0, P_1\}, P_2)$}
 \begin{algorithmic}[1]
    \Require
      $P_0, P_1$ hold the share representation of $z\in \mathbb{Z}/N\mathbb{Z}$ and $a \in \mathbb{F}_2$
    \Ensure
      $P_0$, $P_1$ get the share representation of $az \in \mathbb{Z}/N\mathbb{Z}$.
      \State $P_0$ views $z_0$ as an element in $\mathbb{Z}/2^N\mathbb{Z}$,  $P_1$ view $z_1$ as element in $\mathbb{Z}/2N\mathbb{Z}$;
 \State $P_0, P_1$ run the \textbf{GM}($\{ \pm 1 \}, \mathbb{Z}/2N\mathbb{Z}$) protocol  to get the  share representation $(u_0, u_1)$ of $(-1)^az \in \mathbb{Z}/2N\mathbb{Z}$;
 \State $P_0$ computes $v_0:=\lfloor \frac{x_0-u_0}{2} \rfloor \mod N$, $P_1$ computes $v_1:= \lceil \frac{x_1-u_1}{2}  \rceil \mod N$. Here ``$\lfloor \quad \rfloor$'' means the floor function, and ``$\lceil \quad \rceil$'' means the ceil function; \\
     \Return $(v_0, v_1)$.
  \end{algorithmic}
   \label{alg:SSS_even}
\end{algorithm}

\subsection{Select Share protocol SS(N)}

In this subsection we give our 3-party protocol realizing the functionality $\mathcal{F}_{SS}$  securely in  $\mathcal{F}_{SSS}$-hybrid model as the following Algorithm \ref{alg:SSN}. \begin{algorithm}[htbp]
  \caption{Select Shares \textbf{SS(N)}, $\prod _{SS} (\{P_0, P_1\}, P_2)$}
\begin{algorithmic}[1]
    \Require
      $P_0, P_1$ hold the share representation of $x, y\in \mathbb{Z}/N\mathbb{Z}$ and $a \in \mathbb{F}_2$
    \Ensure
      $P_0$, $P_1$ get the share representation of $a(y-x)+x \in \mathbb{Z}/N\mathbb{Z}$;
      \State $P_0, P_1$ compute $z:=y-x \in \mathbb{Z}/N\mathbb{Z}$;
      \State $P_0, P_1$ run the SSS(N) protocol to compute $v=az \in \mathbb{Z}/N\mathbb{Z}$;
      \State $P_0, P_1$ compute $u=v+x \in \mathbb{Z}/N\mathbb{Z}$ \\
           \Return $(u_0, u_1)$.
  \end{algorithmic}
    \label{alg:SSN}
\end{algorithm}

It easy to see that, the communication of our \textbf{SS(N)} protocol is same as the communication of protocol \textbf{SSS(N)},
which is same as that of \textbf{GM}($\{ \pm 1 \}, \mathbb{Z}/N\mathbb{Z}$) if $N$ is odd, or \textbf{GM}($\{ \pm 1 \}, \mathbb{Z}/2N\mathbb{Z}$) if $N$ is even. Hence. the communication of our \textbf{SS(N)} is shown as in Table \ref{T_all_protocols}.

We compare our \textbf{SS(N)} protocol to the Select Share protocol in SecureNN \cite{secureNN} (Table \ref{T_SSN}), which is
the state of art of Select Share. It easy to see that the online communication, round and the total 
(online+offline) communication of 
our protocols less than the Select Share protocol in SecureNN \cite{secureNN} when $\log N \geq 3$.


\begin{table*}[htbp]
\caption{Comparison to exists  Select Share protocol}
\begin{center}
\begin{tabular}{c|c|c|c|c|c}
\hline
 N & Protocol  & offline com.  & online comm.  & online round & total comm.\\ \hline
 odd N & SS(N)  &   $ \log N$       &     \color{red}{$ 2(1+\log N)$ }        &  \color{red}{1} & \color{red}{$2+3 \log N$}    \\
 even N & SS(N)  &   $\log N+1$       &     \color{red}{$ 2(2+\log N)$}         &  \color{red}{1} & \color{red}{$ 5+3 \log N$}    \\
N &$\prod _{SS}$ in SecureNN \cite{secureNN} & 0 &  $5 \log N$ & $2$ & $5 \log N$  \\ 
\hline 
$2^{32}$ & SS(N) & 33 & \color{red}{68} & \color{red}{1} & \color{red}{101} \\
$2^{32}$ & $\prod _{SS}$ in SecureNN \cite{secureNN} & 0 & 160 & 2 & 160 \\
\hline
$2^{64}$ & SS(N) & 65 & \color{red}{132} & \color{red}{1} & \color{red}{197} \\
$2^{64}$ & $\prod _{SS}$ in SecureNN \cite{secureNN} & 0 & 320 & 2 & 320 \\
\hline
$2^{128}$ & SS(N) & 129 & \color{red}{260} & \color{red}{1} & \color{red}{389} \\
$2^{128}$ & $\prod _{SS}$ in SecureNN \cite{secureNN} & 0 & 640 & 2 & 640 \\
\hline
\label{T_SSN}
\end{tabular}
\end{center}
\end{table*}

\section{MPC protocol for ReLU}
\label{S_ReLU}

In the fixed point representation of real number, we usually use two's complement to represent a negative number, hence to compute the $\mbox{ReLU}(x)$ for a number $x\in \mathbb{Z}/2^n\mathbb{Z}$,
we need to compute
\[
\mbox{ReLU}(x)= \left \{
\begin{array}{ll}
x & \mbox{ if } x< 2^{n-1} \\
0 & \mbox{ otherwise }
\end{array}
\right .
\]
i.e., $\mbox{ReLU}(x)=\mbox{DReLU}(x)x$.

Algorithm \ref{alg:ReLU} gives our 3-party protocol for realizing $\mathcal{F}_{ReLU}$ securely in the  ($\mathcal{F}_{DReLU}, \mathcal{F}_{SSS}$)-hybrid model.

\begin{algorithm*}[htb]
  \caption{\textbf{ReLU(n):}, $\prod _{ReLU} (\{P_0, P_1\}, P_2)$ }
  \label{alg:ReLU}
  \begin{algorithmic}[1]
    \Require
      $P_0, P_1$ hold a share representation $x = u+ v$ in $\mathbb{Z}/ 2^n \mathbb{Z}$
    \Ensure
      $P_0$, $P_1$ get the share representation of $ReLU(x)$ in $\mathbb{Z}/2\mathbb{Z}$;
    \State $P_0, P_1$ call the \textbf{DReLU(n)}  to get the share representation of $DReLU(x)$; 
    \State $P_0, P_1$ call the \textbf{SSS($2^n$)} protocol to get the share representation $(u_0,u_1)$ of $ReLU(x)$;\\
    \Return $(u_0, u_1)$.
  \end{algorithmic}
\end{algorithm*}

The communication and round of our \textbf{ReLU}(n) protocol is equal to the sum of that of  \textbf{DReLU}(n) and \textbf{SSS($2^n$)}.
We show them in the Table \ref{T_all_protocols}.

%

Now let us compare the communication of our $ReLU(n)$ protocol to that in SecureNN (\cite{secureNN}) and CrypTFlow (\cite{CrypTFlow}) in the Table \ref{T_Compare_ReLU} and from now on let $p$ be a prime number with $p\geq n+2$.

\begin{table*}[htbp]
\label{T_Compare_ReLU}
\caption{Comparison to exists  ReLU protocols}
\begin{center}
\begin{threeparttable}
\begin{tabular}{c|c|c|c|c|c}
\hline
 n & Protocol  & offline com.  & online comm.  & online round & total comm.\\ \hline
 n & \color{red}{Our} &   \color{red}{$(2n-1)\log p \tnote{*} \atop +n+2$}       &     \color{red}{$2n \log p + \atop 5n  +  2$}        &  \color{red}{5} & \color{red}{$(4n-1) \log p + \atop  6n  +4$}    \\
n & SecureNN \cite{secureNN}  & 0                                &  $ 8n \log p + 24n$ & 10 & $8n \log p + 24n$ \\
n & CrypTFlow \cite{CrypTFlow} &       $ 0 $  &$6n \log p + 19n$ &  10  & $6n \log p + 19n$ \\
\hline
 32 &\color{red}{Our}    & \color{red}{362.2}  & \color{red}{495.4} & \color{red}{5} & \color{red}{857.6} \\
32 & SecureNN \cite{secureNN}  & 0 &  2101.6 & 10 & 2101.6 \\
32 & CrypTFlow \cite{CrypTFlow} &       $ 0 $  & 1608.3 &  10  & 1608.3 \\
\hline
 64 & \color{red}{Our}    & \color{red}{836.4}  & \color{red}{1098.5} & \color{red}{5} & \color{red}{1934.9}\\
64 & SecureNN \cite{secureNN}  & 0 &  4641.8 & 10 & 4641.8 \\
64 & CrypTFlow \cite{CrypTFlow} &       $ 0 $  & 3545.4 &  10  & 3545.4 \\
\hline
 128 & \color{red}{Our}    & \color{red}{1923.5}  & \color{red}{2442.6} & \color{red}{5} & \color{red}{4366.1}\\
 128 & SecureNN \cite{secureNN}  & 0 &  10274.2 & 10 & 10274.2 \\
 128 & CrypTFlow \cite{CrypTFlow} &       $ 0 $  & 7833.7 &  10  & 7833.7 \\
\hline

\end{tabular}
\begin{tablenotes}
        \footnotesize
        \item[*]  $p\geq n+2$ is a prime number. 
      \end{tablenotes}
    \end{threeparttable}
\end{center}
\end{table*}

\begin{table}
\begin{threeparttable}
\begin{tabular}{c|c|c}
\hline
 Protocol  & correct range  & action of third party \\ \hline
 \color{red}{Our} &  all & receive only \\
 SecureNN \cite{secureNN} & $|x|<2^k  \tnote{*} $ & receive and send \\
  CrypTFlow \cite{CrypTFlow} & $|x|<2^k  \tnote{*} $ & receive and send \\
  \hline
\end{tabular}
\begin{tablenotes}
        \footnotesize
        \item[*]  $k<n$. 
      \end{tablenotes}
    \end{threeparttable}
\end{table}

\section{Conclusion and Future work}
In this paper, we defined three new functionality for the mathematical object: G-module.  And gave 3-party
secure protocols realizing theses functionality.
As the applications of the protocols of G-module, we gave new 3-party protocol securely realizing the functionalities 
Select Share, DReLU, and ReLU. Our new protocols are better than the state of art in communication, correctness and security.

In the future, we will  construct a system for secure deep leaning with our protocols ReLU and DReLU.
We also  will use the tool G-module to improve more protocols in MPC.

\clearpage


\end{document}